\documentclass[twocolumn,prl]{revtex4}

\usepackage{amsfonts}
\usepackage{amsthm}
\usepackage{amssymb}
\usepackage{amsmath}

\newtheorem{thm}{Theorem}
\newtheorem{prop}[thm]{Proposition}

\newtheorem{lem}[thm]{Lemma}

\newcommand{\be}{\begin{equation}}
\newcommand{\ee}{\end{equation}}

\newcommand{\W}{\mathcal{W}_{\lambda,d}}
\newcommand{\D}{\mathcal{D}_{\lambda,d}}

\newcommand{\C}{\mathbb{C}}

\newcommand{\ket}[1]{|#1\rangle}
\newcommand{\braket}[2]{\langle #1|#2\rangle}
\newcommand{\ketbra}[2]{|#1\rangle\langle #2|}
\newcommand{\pure}[1]{\ketbra{#1}{#1}}
\newcommand{\Tr}{\mathop{\mathrm{Tr}}}
\providecommand{\one}{\leavevmode\hbox{\small1\kern-3.8pt\normalsize1}}

\begin{document}

\title{Multiplicativity of the maximal output $2$-norm for depolarized Werner-Holevo channels.}
\author{S. Michalakis}
\email{spiros@math.ucdavis.edu}
\affiliation{Department of Mathematics, University of California at Davis - Davis CA 95616, USA}

\pacs{05.50.+q}

\begin{abstract}

We study the multiplicativity of the output $2$-norm for depolarized Werner-Holevo channels and show
that multiplicativity holds for a product of two identical channels in this class. Moreover, it shown that the depolarized Werner-Holevo channels do not satisfy the entrywise positivity condition introduced by C. King and M.B. Ruskai, which suggests that the main result is non-trivial.

\end{abstract}

\maketitle

%%%%%%%%%%%%%%%%%%%%%%%%%%%%%%%%
%                                           				       			           %
%       				 Introduction                       			  %
%                                                                                        			  %
%%%%%%%%%%%%%%%%%%%%%%%%%%%%%%%%

\section{The setup and main result} 
The $d$-dimensional Werner-Holevo channel ${\cal W}_d(\rho) = \frac{1}{d-1} ((\Tr(\rho)) \one_d - \rho^T)$ is known~\cite{WH} to give a counterexample to the multiplicativity of the maximal output $p$-norm for $p > 4.79$, when $d=3$.
Nevertheless, it has been shown~\cite{Datta, AF} that ${\cal W}_d(\rho)$ satisfies multiplicativity for $1\leq p \leq 2$.
It is natural then to study the output $p$-norm of depolarized Werner-Holevo channels 
$$\W(\rho) = \lambda \rho + (1-\lambda) {\cal W}_d(\rho),$$ and ask if those channels
satisfy multiplicativity for $p$-norms with $p \leq 2$.

We focus our attention to the study of the output $2$-norm for the tensor product channel $\W\otimes\W$
acting on bipartite states in $M_d(\C)\otimes M_d(\C)$ and show that multiplicativity is satisfied for this norm.

A direct computation of the eigenvalues of $\W\otimes\W(\pure{\psi_{12}})$ turns out to be much harder for $0 < \lambda < 1$, than for the boundary cases $\lambda = 0, 1$. The reason is that the
output consists of a combination of the input state and its transpose/partial transpose, which in
general do not share a common eigenbasis. To work around this difficulty, we compute explicitly the
output $2$-norm of $\W\otimes\W(\pure{\psi_{12}})$ and the maximal output $2$-norm of $\W$ and study the difference 
\begin{equation}\label{eqn:output_dist}
\D(\psi_{12}) = (\| \W \|_2^2)^2 - \|\W \otimes \W(\pure{\psi_{12}}) \|_2^2
\end{equation}
We show that $\D \ge 0$ for all input states and $\lambda \in [0,1],\,d\ge 2$.
We begin with the computation of $\| \W \|_2^2$ in the following Lemma.

\begin{lem}\label{lem:max_norm}
The (squared) maximal output $2$-norm of $\W$ is given by
$$\|\W\|_2^2 = \frac{(d-2)\lambda^2 + 1}{d-1}$$
\end{lem}
\begin{proof}
It is easy to check that $\|\W(\pure{\psi})\|_2^2$ is
\begin{eqnarray*}
&=& \Tr(\W(\pure{\psi})^2) \\
&=& \lambda^2 + \frac{2\lambda(1-\lambda)(1-|\braket{\psi}{\overline{\psi}}|^2)}{d-1} + \frac{(1-\lambda)^2}{d-1}\\
&\le& \frac{(d-2)\lambda^2 +1}{d-1},
\end{eqnarray*}
where $\ket{\overline{\psi}}$ denotes the complex conjugate of $\ket{\psi}$ in the standard basis.
Taking $\ket{\psi} = \frac{\ket{0} + i\ket{1}}{\sqrt{2}}$, with $\ket{0},\ket{1}$ two standard basis vectors, we see that equality can be achieved in the above expression and the result follows.
\end{proof}

We now turn our attention to the more complicated output $2$-norm of $\W\otimes\W(\pure{\psi_{12}})$.

\begin{lem}\label{lem:output_norm}
The (squared) output $2$-norm $\|\W\otimes\W(\pure{\psi_{12}})\|_2^2$ is given by:
\begin{eqnarray*}
& &(\|\W\|_2^2)^2\\ 
&+& S_\lambda^2 \, |\braket{\psi_{12}}{\overline{\psi_{12}}}|^2\\
&-& 2 (S_\lambda + R_\lambda^2)
(S_\lambda + (d-2)Q_\lambda^2\Big)
\big(1-\|\rho_1\|_2^2\big)\\
&-& S_\lambda \|\W\|_2^2 \Tr(\rho_1\rho_1^T + \rho_2\rho_2^T),
\end{eqnarray*}
where $Q_\lambda = \frac{1-\lambda}{d-1}$, $R_\lambda = \lambda - Q_\lambda$, $S_\lambda = 2\lambda Q_\lambda$, $\rho_1 = {\Tr}_2\pure{\psi_{12}}$, $\rho_2 = {\Tr}_1\pure{\psi_{12}}$ and $T$ denotes
transposition.
\end{lem}
\begin{proof}
It is easy to check that 
\begin{eqnarray*}
\W\otimes\W(\pure{\psi_{12}}) &=& \lambda^2 \pure{\psi_{12}}\\
&+& Q_\lambda R_\lambda [\rho_1\otimes\one_d + \one_d\otimes\rho_2]\\
&+& Q_\lambda^2\big[\one_d\otimes\one_d + \pure{\overline{\psi_{12}}}\big]\\
&-& \frac{S_\lambda}{2}(\pure{\psi_{12}})^{T_1}\\
&-& \frac{S_\lambda}{2}(\pure{\psi_{12}})^{T_2},
\end{eqnarray*}
where $T_1, T_2$ denote partial transposition w.r.t. the $1^{\rm st}$, $2^{\rm nd}$ tensor factor, respectively.
Taking the trace after squaring the above expression and noting that 
$\Tr \pure{\psi_{12}}(\pure{\psi_{12}})^{T_k} = \Tr \rho_k \rho_k^T$, for $k=1,2$ (which one can show using the Schmidt decomposition of $\ket{\psi_{12}}$), we get the desired result.
\end{proof}

The following general inequality will be very useful in the proof of the main theorem, so we state it
here as a lemma.
\begin{lem}\label{ineq:two_norm}
Let $\sigma_1 \le \sigma_2 \le \ldots \le \sigma_d$ be non-negative numbers that sum up to $1$. Then, the following inequality holds:
\begin{equation*} \sigma_d \ge \sum_{\alpha = 1}^d \sigma_\alpha^2
\end{equation*}
\end{lem}
\begin{proof}
The r.h.s. of the inequality can be thought of as the expected value of the random variable $X$ given by $\mathrm{Pr}(X = \sigma_{\alpha}) = \sigma_\alpha$. The upper bound then follows immediately.
\end{proof}

%%%%%%%%%%%%%%%%%%%%%%%%%%%%%%%%
%                                            				       			  %
%                                    Proof of Main Theorem          			  %
%                                                                                        			  %
%%%%%%%%%%%%%%%%%%%%%%%%%%%%%%%%
\section{Proof of the Main Result}
In this section, we will show that the difference $\D$ defined in~(\ref{eqn:output_dist}) is always non-negative, which is equivalent to multiplicativity of the output $2$-norm for $\W$. We state this as a theorem:
\begin{thm}
For the depolarized Werner-Holevo channel $\W$, we have for $\lambda \in [0,1], d\ge 2$:
$$\|\W\otimes\W\|_2 = \|\W\|_2^2$$
\end{thm}

\begin{proof}
From Lemma~\ref{lem:output_norm} we see that the condition $\D(\ket{\psi_{12}}) \ge 0$ is equivalent to
\begin{eqnarray*}
S_\lambda^2\, 
|\braket{\psi_{12}}{\overline{\psi_{12}}}|^2 &\le&
2 (S_\lambda^2 + P_\lambda^2)
\big(1-\|\rho_1\|_2^2\big)\\
&+& S_\lambda \|\W\|_2^2 \Tr(\rho_1\rho_1^T + \rho_2\rho_2^T),
\end{eqnarray*}
where $P_\lambda^2 = [Q_\lambda^2+(d-2)R_\lambda^2]S_\lambda
+ (d-2)Q_\lambda^2 R_\lambda^2 \ge 0$.
Using Lemma~\ref{lem:max_norm} to write $\|\W\|_2^2$ as $(1+\sqrt{d-1})S_\lambda + (\lambda - \frac{1-\lambda}{\sqrt{d-1}})^2$, we see that it is sufficient to prove the following inequality\begin{equation}\label{ineq:output_dist}
|\braket{\psi_{12}}{\overline{\psi_{12}}}|^2 \le
2\big(1-\|\rho_1\|_2^2\big) + (1+\sqrt{d-1}) \Tr(\rho_1\rho_1^T + \rho_2\rho_2^T)
\end{equation}
(the boundary cases $\lambda = 0,1$ follow from $1 \ge \|\rho_1\|_2^2$).

We will now make use of the Schmidt decomposition of the input state $\pure{\psi_{12}}$, given by
$\ket{\psi_{12}} = \sum_{\alpha} \sqrt{\sigma_{\alpha}} \ket{\alpha_1}\otimes \ket{\alpha_2}$, where 
$\{ \ket{\alpha_1}\},\{ \ket{\alpha_2}\}$ are orthonormal sets in $M_d(\C)$.
We have that $\|\rho_1\|_2^2 = \sum_{\alpha = 1}^d \sigma_\alpha^2$, where some of the $\sigma_\alpha$ may be zero. Applying Lemma~\ref{ineq:two_norm} (and borrowing its notation w.l.o.g.), 
it follows that $\|\rho_1\|_2^2 \le \sigma_d$. Moreover, it becomes clear now that in order to prove~(\ref{ineq:output_dist}), it is sufficient to show:
\begin{equation}\label{ineq:output_dist2}
|\braket{\psi_{12}}{\overline{\psi_{12}}}|^2 \le
2\big(1-\sigma_d) + (1+\sqrt{d-1}) \Tr(\rho_1\rho_1^T + \rho_2\rho_2^T)
\end{equation}
for $\sigma_d \ge 1/2$, since $|\braket{\psi_{12}}{\overline{\psi_{12}}}| \le 1$ and 
$\Tr(\rho_1\rho_1^T + \rho_2\rho_2^T) \ge 0$.
We now use the triangle inequality to get an estimate for the l.h.s. of~(\ref{ineq:output_dist2}),
\begin{eqnarray}\label{eqn:inner_transpose}
|\braket{\psi_{12}}{\overline{\psi_{12}}}| &=& |\sum_{\alpha,\beta} \sqrt{\sigma_\alpha \sigma_\beta}
\braket{\alpha_1}{\overline{\beta_1}}\braket{\alpha_2}{\overline{\beta_2}}|\nonumber\\
&\le& \sum_{\alpha,\beta} \sqrt{\sigma_\alpha \sigma_\beta}
|\braket{\alpha_1}{\overline{\beta_1}}||\braket{\alpha_2}{\overline{\beta_2}}|
\end{eqnarray}
We will need to treat dimensions $d \le 4$ and $d \ge 5$ separately.
For $d \le 4$ we use Cauchy-Schwarz to get the following estimate for
$$\Big(\sum_{\alpha,\beta} \sqrt{\sigma_\alpha \sigma_\beta}
|\braket{\alpha_1}{\overline{\beta_1}}||\braket{\alpha_2}{\overline{\beta_2}}|\Big)^2
$$
\begin{eqnarray}
&\le& \Big(\sum_{\alpha,\beta} \sigma_\alpha \sigma_\beta
|\braket{\alpha_1}{\overline{\beta_1}}|^2\Big) 
\Big( \sum_{\alpha,\beta} |\braket{\alpha_2}{\overline{\beta_2}}|^2\Big) \nonumber \\
&\le& d \sum_{\alpha,\beta} \sigma_\alpha \sigma_\beta
|\braket{\alpha_1}{\overline{\beta_1}}|^2 \label{cs1}
\end{eqnarray}
where we have used Parseval's identity in the last inequality. The same inequality is, of course,
true for the second tensor factor. Using~(\ref{ineq:output_dist2}) and~(\ref{eqn:inner_transpose}) along with the fact that
\begin{equation}
\Tr(\rho_k \rho_k^T) =  \sum_{\alpha,\beta} \sigma_\alpha \sigma_\beta
|\braket{\alpha_k}{\overline{\beta_k}}|^2\qquad k=1,2
\end{equation}
we see from estimate~(\ref{cs1}) that it is sufficient to show that $d \leq 2(1+\sqrt{d-1})$, which is true for $d \le 4$.
We now turn our attention to the case $d \ge 5$.
We will need a different estimate than the one given in~(\ref{cs1}), since we need to make use
of the assumption that $\sigma_d \ge 1/2$ in order to lower the factor $d$ in~(\ref{cs1}).
We start by using Cauchy-Schwarz to get the following upper bound:
\begin{equation}\label{cs2}
\Big(\sum_{\alpha,\beta} \sqrt{\sigma_\alpha \sigma_\beta}
|\braket{\alpha_1}{\overline{\beta_1}}||\braket{\alpha_2}{\overline{\beta_2}}|\Big)^2
\le 3 (I_1^2 + I_2^2 + I_3^2),
\end{equation}
where 
\begin{eqnarray*}
I_1 &=& \sum_{\alpha = d,\beta} \sqrt{\sigma_\alpha \sigma_\beta}
|\braket{\alpha_1}{\overline{\beta_1}}||\braket{\alpha_2}{\overline{\beta_2}}| \\
I_2 &=& \sum_{\alpha \ne d ,\beta = d} \sqrt{\sigma_\alpha \sigma_\beta}
|\braket{\alpha_1}{\overline{\beta_1}}||\braket{\alpha_2}{\overline{\beta_2}}| \\
I_3 &=& \sum_{\alpha \ne d, \beta \ne d} \sqrt{\sigma_\alpha \sigma_\beta}
|\braket{\alpha_1}{\overline{\beta_1}}||\braket{\alpha_2}{\overline{\beta_2}}|
\end{eqnarray*}
A further application of Cauchy-Schwarz on $I_1, I_2, I_3$ will give us the desired result.
We start with an estimate for $I_1$, since $I_2$ is very similar.
Noting that one of the summation indices is fixed to $d$, we get
\begin{eqnarray*}
I_1^2 &=& \Big(\sum_{\alpha = d,\beta} \sqrt{\sigma_\alpha \sigma_\beta}
|\braket{\alpha_1}{\overline{\beta_1}}||\braket{\alpha_2}{\overline{\beta_2}}|\Big)^2 \\
&\le& \Big(\sum_{\alpha = d,\beta} \sigma_\alpha \sigma_\beta
|\braket{\alpha_1}{\overline{\beta_1}}|^2 \Big)\Big(\sum_{\alpha = d,\beta} |\braket{\alpha_2}{\overline{\beta_2}}|^2\Big)\\
&\le& \sum_{\alpha = d,\beta} \sigma_\alpha \sigma_\beta
|\braket{\alpha_1}{\overline{\beta_1}}|^2\\
&\le& \Tr(\rho_1 \rho_1^T)
\end{eqnarray*}
Similarly, we see that $I_2^2 \le \Tr(\rho_2 \rho_2^T)$. Since $1+\sqrt{d-1} \ge 3$ for $d \ge 5$, we
see from~(\ref{ineq:output_dist2}) and~(\ref{cs2}) that it remains to show $3\, I_3^2 \le 2\,(1-\sigma_d)$.
We have
\begin{eqnarray*}
I_3^2 &=& \Big(\sum_{\alpha \ne d,\beta \ne d} \sqrt{\sigma_\alpha \sigma_\beta}
|\braket{\alpha_1}{\overline{\beta_1}}||\braket{\alpha_2}{\overline{\beta_2}}|\Big)^2 \\
&\le& \Big(\sum_{\alpha \ne d,\beta \ne d} \sigma_\alpha
|\braket{\alpha_1}{\overline{\beta_1}}|^2 \Big)
\Big(\sum_{\alpha \ne d,\beta \ne d} \sigma_\beta |\braket{\alpha_2}{\overline{\beta_2}}|^2\Big)\\
&\le& \Big(\sum_{\alpha \ne d} \sigma_\alpha \Big)^2\\
&=& (1-\sigma_d)^2
\end{eqnarray*}
It remains to show that $3\, (1-\sigma_d)^2 \le 2 \, (1-\sigma_d) 
\Leftrightarrow (1-\sigma_d)(3\sigma_d -1) \ge 0$, which follows from our assumption that $\sigma_d\in[\frac{1}{2},1]$.
\end{proof}

\section{Discussion}
We have shown that for depolarized Werner-Holevo channels the maximum output $2$-norm is
multiplicative. For $\lambda \in (0,1)$ and $d \ge 3$, the depolarized Werner-Holevo maps do not satisfy the entrywise-positivity (EP) condition introduced by C. King and M.B. Ruskai in~\cite{KR1, KR2}. 
This suggests that some elements of the above proof may be useful when tackling
the multiplicativity of the maximal output $2$-norm for arbitrary channels.

\begin{prop}
The depolarized Werner-Holevo channels $\W$ with $\lambda \in (0,1), \, d \ge 3$ do not satisfy the entrywise-positivity (EP) condition:
\begin{equation*}
\Tr \W(\ketbra{e_l}{e_i})\W(\ketbra{e_j}{e_k}) \ge 0, \qquad \forall i,j,k,l,
\end{equation*}
and $\{\ket{e_i}\}_{i=1}^d$ some orthonormal basis of $\C^d$.
\end{prop}
\begin{proof}
One can check that $\Tr \W(\ketbra{e_l}{e_i})\W(\ketbra{e_j}{e_k})$ is given by:
\begin{eqnarray*}
& & 
\Big[\lambda^2+\Big(\frac{1-\lambda}{d-1}\Big)^2\Big] \delta_{i,j}\delta_{k,l}\\
&+& \Big[\frac{2\lambda(1-\lambda)}{d-1}+(d-2)\Big(\frac{1-\lambda}{d-1}\Big)^2\Big] \delta_{i,l}\delta_{j,k}\\
&-& \frac{2\lambda(1-\lambda)}{d-1} \braket{e_i}{\overline{e_k}}\braket{\overline{e_l}}{e_j}
\end{eqnarray*}
where $\ket{\overline{e_k}}$ denotes the complex conjugate of $\ket{e_k}$, as before.
Now, taking $i=j, k\ne l$ in the above expression, we see that the EP condition implies:
$$ \braket{e_i}{\overline{e_k}}\braket{\overline{e_l}}{e_i} \le 0, \qquad \forall i, k\ne l.$$
Summing over $i$ in the above inequality gives us $0$, which implies that:
\be\label{cond}
\braket{e_i}{\overline{e_k}}\braket{\overline{e_l}}{e_i} = 0, \qquad \forall i, k\ne l.
\ee
Fixing $l$, we choose $i = \pi(l)$ such that $\braket{\overline{e_l}}{e_{\pi(l)}} \ne 0$ (we can always find such a $\pi(l)$, since otherwise $\ket{\overline{e_l}} = 0$; a contradiction to $\ket{\overline{e_l}}$ being an orthonormal basis vector). 
Condition~(\ref{cond}) then implies that $\braket{e_{\pi(l)}}{\overline{e_k}} = 0, \, \forall k\ne l$. 
Since the $\{\ket{\overline{e_k}}\}$ form an orthonormal basis, it follows that 
$\ket{e_{\pi(l)}} = \ket{\overline{e_l}}, \forall l$.
We may now rewrite the EP condition as:
\begin{eqnarray*}
& & \Big[\lambda^2+\Big(\frac{1-\lambda}{d-1}\Big)^2\Big] \delta_{i,j}\delta_{k,l}\\
&+& \Big[\frac{2\lambda(1-\lambda)}{d-1}+(d-2)\Big(\frac{1-\lambda}{d-1}\Big)^2\Big] \delta_{i,l}\delta_{j,k}\\
&\ge& \frac{2\lambda(1-\lambda)}{d-1} \delta_{i, \pi(k)}\delta_{j, \pi(l)}
\end{eqnarray*}
Choosing $i = \pi(k), j = \pi(l)$ and $k \ne l$, the above condition becomes:
\begin{equation*}
\Big[\frac{2\lambda(1-\lambda)}{d-1}+(d-2)\Big(\frac{1-\lambda}{d-1}\Big)^2\Big] \delta_{\pi(k),l}\delta_{\pi(l),k}
\ge \frac{2\lambda(1-\lambda)}{d-1}
\end{equation*}
The EP condition forces $\pi(l) = k, \forall k \ne l$ (note that $\pi(k) = l$ then follows from the definition of $\pi(k)$,) which is impossible for $d \ge 3$. For $d=2$, choosing $\ket{e_1} = \frac{\ket{0} + i \ket{1}}{\sqrt{2}}$ satisfies the EP condition.
\end{proof}

\begin{acknowledgments}
The author would like to thank his advisor Bruno Nachtergaele for his constant support, Mary Beth Ruskai for suggesting the problem and her many helpful discussions and Koenraad Audenaert for testing numerically (and disproving) one of the conjectures that came up along the way.
Support from NSF Grant DMS-0605342 is acknowledged.
\end{acknowledgments}


\begin{thebibliography}{0}
\bibitem{WH}
R. F. Werner and A. S. Holevo, ``Counterexample to an additivity conjecture for output purity
of quantum channels",
J. Math. Phys. {\bf 43}, 4353-4357 (2002).

\bibitem{Datta}
N. Datta, ``Multiplicativity of maximal $p$-norms in Werner-Holevo channels for $1 \le p \le 2$",
arXiv:quant-ph/0410063v1.

\bibitem{AF}
R. Alicki and M. Fannes, ``Note on multiple additivity of minimal Renyi entropy output of the
Werner-Holevo channels'',
arXiv:quant-ph/0407033.

\bibitem{KR1}
C. King and M.B. Ruskai, ``Multiplicativity properties of entrywise positive maps'', 
arXiv:quant-ph/0409181v2.

\bibitem{KR2}
C. King and M.B. Ruskai, ``Comments on multiplicativity of maximal p-norms when $p=2$'', 
arXiv:quant-ph/0401026v1.


\end{thebibliography}
\end{document}